\documentclass[aps,pra,twocolumn,amsmath,amssymb,showpacs,superscriptaddress]{revtex4}

\newcommand{\bra}[1]{\langle#1|}
\newcommand{\ket}[1]{|#1\rangle}

\usepackage[dvips]{graphicx}
\usepackage{graphics}
\usepackage{mathrsfs}

\begin{document}

\bibliographystyle{apsrev}

\title{Practical effects in the preparation of cluster states using weak non-linearities}

\author{Peter P. Rohde}
\email[]{rohde@physics.uq.edu.au}
\homepage{http://www.physics.uq.edu.au/people/rohde/}
\affiliation{Centre for Quantum Computer Technology, Department of Physics\\ University of Queensland, Brisbane, QLD 4072, Australia}

\author{William J. Munro}
\affiliation{National Institute of Informatics, 2-1-2 Hitotsubashi, Chiyoda-ku, Tokyo 101-8430, Japan}
\affiliation{Hewlett-Packard Laboratories, Filton Road, Stoke Gifford, Bristol BS34 8QZ, United Kingdom}

\author{Timothy C. Ralph}
\affiliation{Centre for Quantum Computer Technology, Department of Physics\\ University of Queensland, Brisbane, QLD 4072, Australia}

\author{Peter van Loock}
\affiliation{National Institute of Informatics, 2-1-2 Hitotsubashi, Chiyoda-ku, Tokyo 101-8430, Japan}

\author{Kae Nemoto}
\affiliation{National Institute of Informatics, 2-1-2 Hitotsubashi, Chiyoda-ku, Tokyo 101-8430, Japan}

\date{\today}

\frenchspacing

\begin{abstract}
We discuss experimental effects in the implementation of a recent scheme for performing bus mediated entangling operations between qubits. Here a bus mode, a strong coherent state, successively undergoes weak Kerr-type non-linear interactions with qubits. A quadrature measurement on the bus then projects the qubits into an entangled state. This approach has the benefit that entangling gates are non-destructive, may be performed non-locally, and there is no need for efficient single photon detection. In this paper we examine practical issues affecting its experimental implementation. In particular, we analyze the effects of post-selection errors, qubit loss, bus loss, mismatched coupling rates and mode-mismatch. We derive error models for these effects and relate them to realistic fault-tolerant thresholds, providing insight into realistic experimental requirements.
\end{abstract}


\maketitle

\section{Introduction}
Quantum optical systems have received a lot of attention as a potential candidate for the implementation of scalable quantum computation. In particular, linear optics quantum computing (LOQC) has shown that it is possible to implement scalable quantum computation using just single photon sources, photo-detectors and passive linear optics \cite{bib:KLM01, bib:Kok07, bib:Ralph06}. LOQC suffers from the fact the entangling gate operations are inherently non-deterministic. Knill, Laflamme and Milburn \cite{bib:KLM01} showed that this could be overcome using an elaborate and complicated scheme based on encoding and gate teleportation \cite{bib:GottesmanChuang99}. Although this works in principle, the scaling of physical resource requirements are extremely large. Variations of LOQC based on the cluster-state model \cite{bib:Nielsen04,bib:BrowneRudolph05} and parity encoding \cite{bib:RalphHayes05} have significantly reduced physical resource requirements. This reduction in physical resources stems from the fact that these approaches take advantage of offline state preparation. This allows, for example, the preparation of a large cluster state to be broken down into the preparation of many smaller cluster states, which succeed with higher probability. This approach has allowed physical resource to be reduced by several orders of magnitude. Nonetheless, LOQC remains technologically challenging due to the difficulty in preparing indistinguishable, triggered single photon states and performing high-efficiency photo-detection.

An alternate `hybrid' approach has recently been proposed where qubits (which may be either photonic or solid-state) interact indirectly via a strong coherent optical beam, which acts as a `bus' \cite{bib:Munro05, bib:Spiller06, bib:BarrettMilburn06}. Entangling gates are implemented through successive interactions between qubits and the bus, followed by a projective quadrature measurement on the bus. This approach has several inherent benefits. First, there is no need for single-photon detection. Instead, efficient homodyne detection may be employed when measuring the bus mode. Second, the qubits may be spatially separated since they do not interact directly. For this reason this approach has also been suggested for entanglement distribution \cite{bib:Loock06,bib:Ladd06}. Third, this approach is applicable to both photonic and solid-state systems. There are several different types of bus-mediated gate, including a controlled-phase gate and several implementations of the parity gate.

In this paper we examine effects that are likely to be significant in the practical implementation of such a scheme. We specifically focus on the `parity gate' \cite{bib:Munro05}, the simplest bus-mediated gate, which projects a two qubit state into the even or odd parity subspace. This gate is of particular relevance since it can be used in the construction of cluster-states \cite{bib:BrowneRudolph05, bib:Louis06}, which are sufficient for universal quantum computation. For this gate we derive error models describing various practical effects, which we relate to estimated fault-tolerant thresholds. This provides insight into realistic technological requirements for the experimental implementation of such a scheme.

This paper is structured as follows. In Section~\ref{sec:background} we review the bus-mediated parity gate. In Section~\ref{sec:post_selection_errors} we discuss the effects of post-selection errors, which are introduced during post-selective measurement of the bus mode, while in Section~\ref{sec:bus_loss} we examine the effects of bus loss. Both of these results were previously known, but we include them here for completeness. In Section~\ref{sec:mis_coupling} we consider the effect of mismatched coupling rates between the two qubit/bus interactions. In Section~\ref{sec:mode_mismatch} we examine the effects of mode-mismatch, which is specific to implementations employing photonic qubits. In Seciton~\ref{sec:self_kerr} we consider the effects of self Kerr effects during the non-linear interactions. We conclude in Section~\ref{sec:conclusion}.

\section{Background} \label{sec:background}
We begin by reviewing the bus-mediated two qubit parity gate. This gate projects an incident state into the even parity subspace with a maximum success probability of 50\%. Consider the completely general two qubit state
\begin{equation} \label{eq:input_state}
\ket{\psi_\mathrm{in}} = \ket\alpha(c_{00}\ket{00} + c_{01}\ket{01} + c_{10}\ket{10} + c_{11}\ket{11}),
\end{equation}
where the first term represents a coherent probe beam that will later be used to mediate interactions between the two qubits. In all our calculations we will assume real $\alpha$ for simplicity. Next consider a non-linear interaction acting between a single qubit and the coherent probe. For photonic qubits this interaction takes the form of a weak cross-Kerr interaction, described by an interaction Hamiltonian of the form
\begin{equation}
\hat{H} = \hbar \chi \hat{n}_q \hat{n}_p,
\end{equation}
where $\hat{n}_q$ and $\hat{n}_p$ are the photon number operators for the qubit and probe modes respectively. Here we have assumed the Hamiltonian is flat in frequency. For an analysis with frequency dependent Hamiltonian, see Ref. \cite{bib:Shapiro07}. Alternately, for solid-state qubits we assume an analogous interaction of the form
\begin{equation}
\hat{H} = \hbar \chi \hat{Z}_q \hat{n}_p,
\end{equation}
where $\hat{Z}_q$ is the Pauli phase-flip operator acting on the qubit. For both of these interactions the unitary operation describing to the interaction takes the form
\begin{eqnarray}
\hat{U}(\theta)\ket\alpha\ket{0} &=& \ket\alpha\ket{0} \nonumber\\
\hat{U}(\theta)\ket\alpha\ket{1} &=& \ket{\alpha e^{i\theta}}\ket{1},
\end{eqnarray}
where $\theta=\chi t$ denotes the coupling strength of the interaction and $t$ is the interaction time. We apply the sequence of operations $\hat{U}_{PB}(-\theta)\hat{U}_{PA}(\theta)\ket{\psi_\mathrm{in}}$ to the input state from Eq.~\ref{eq:input_state}, where $A$ and $B$ denote the two qubits and $P$ denotes the coherent probe beam. This gives the output state,
\begin{eqnarray} \label{eq:output_ideal}
\ket{\psi_\mathrm{out}} &=& c_{00}\ket\alpha\ket{00} + c_{01}\ket{\alpha e^{i\theta}}\ket{01}\nonumber\\
&+& c_{10}\ket{\alpha e^{-i\theta}}\ket{10} + c_{11}\ket\alpha\ket{11}.
\end{eqnarray}
For sufficiently large $\alpha$ and $\theta$, the $\ket\alpha$ and $\ket{\alpha e^{\pm i\theta}}$ components of the probe beam are classically distinguishable in phase space. We perform a $p$-quadrature measurement, conditioning on a window around $p\approx 0$. In this case the probability of detecting the $\ket{\alpha e^{\pm i\theta}}$ components of the probe drop to approximately zero, leaving us with the conditional state
\begin{equation}
\ket{\psi_\mathrm{cond}} = c_{00}\ket{00} + c_{11}\ket{11},
\end{equation}
the even parity projected state, which is described by the even parity projection operator
\begin{equation}
\hat\Lambda_+ = \ket{00}\bra{00} + \ket{11}\bra{11}.
\end{equation}
Realistically only very weak interactions of the above type are experimentally attainable, i.e. in practise $\theta\ll 1$. The crucial feature of this scheme is that a strong coherent probe `amplifies' the effect of the weak coupling. Thus, in principle, even for very small $\theta$ the states $\ket\alpha$ and $\ket{\alpha e^{\pm i\theta}}$ are classically distinguishable for sufficiently large $\alpha$.

While in principle this scheme can work for aribtrary $\alpha$, it is simplest to consider real $\alpha$. If complex $\alpha$ are used, the homodyne measurement must be performed along a rotated axis. Thus, for simplicity we will assume real $\alpha$ in our simulations, which enables us to use straightforward $p$ quadrature measurements.

\section{Post-selection errors} \label{sec:post_selection_errors}
The first source of error we consider is post-selection error. This arises because the different components of the coherent probe beam, $\ket\alpha$ and $\ket{\alpha e^{\pm i\theta}}$, have small, but non-zero overlap. Thus, when applying the $p$-quadrature projection there is some probability that we actually detected a contribution from the $\ket{\alpha e^{\pm i\theta}}$ terms rather than the expected $\ket\alpha$ term. Note that this type of error is intrinsic to the scheme, and is not caused by any type of experimental imperfection. The following result is already known \cite{bib:Munro05}, but we include it for completeness.

Consider the output state given by Eq.~\ref{eq:output_ideal}. Next suppose we post-select onto the $p$-quadrature and detect $p$. We define
\begin{equation}
\gamma_{p,\theta} = \langle p | \alpha e^{i\theta} \rangle = \mathrm{exp}[-\alpha^2 - (p-2i\alpha e^{i\theta})^2/4]
\end{equation}
to be the overlap between position eigenstate $\ket{p}$ and coherent state $\ket{\alpha e^{i\theta}}$.

If we no longer make the approximation that the $\ket{\alpha e^{\pm i\theta}}$ components of the probe beam are never detected, it is evident that upon post-selection of the output state given by Eq.~\ref{eq:output_ideal}, we obtain
\begin{eqnarray}
\ket{\psi_\mathrm{cond}} &=& \gamma_{p,0} c_{00}\ket{00} + \gamma_{p,\theta} c_{01}\ket{01} \nonumber\\
&+& \gamma_{p,-\theta} c_{10}\ket{10} + \gamma_{p,0} c_{11}\ket{11}.
\end{eqnarray}
This corresponds to the measurement projection operator
\begin{equation} \label{eq:ps_error_proj}
\hat\Lambda = \gamma_{p,0}(\ket{00}\bra{00} + \ket{11}\bra{11}) + \gamma_{p,\theta}\ket{01}\bra{01} + \gamma_{p,-\theta}\ket{10}\bra{10},
\end{equation}
which clearly contains undesired odd-parity terms. Consider the behavior of this projector in the limiting cases. First, as $\theta\to 0$, $\hat\Lambda\to\openone$. This is expected, since in this limit there is no phase-space separation between the different parity terms, so we always perform the identity operation. On the other hand, for $\alpha\gg 1$ and $\theta>0$, we have $\gamma_{p\approx 0,\theta}\to 0$ so $\hat\Lambda \to \ket{00}\bra{00} + \ket{11}\bra{11}$, the ideal case.

Eq.~\ref{eq:ps_error_proj} has been grouped such that the first term corresponds to the ideal even parity projector, while the second and third terms correspond to undesired odd parity terms. This grouping allows us to easily relate this projector to the measurement error rate. Namely, the measurement error probability is given by the relative magnitude of the undesired terms,
\begin{equation} \label{eq:post_selection_error}
P_\mathrm{error} = \frac{\left|\gamma_{p,\theta}\right|^2 + \left| \gamma_{p,-\theta}\right|^2}{2 \left|\gamma_{p,0}\right|^2 +\left|\gamma_{p,\theta}\right|^2 + \left|\gamma_{p,-\theta}\right|^2} = \frac{e^{-d^2}}{1+e^{-d^2}},
\end{equation}
where $d = 2 \alpha\,\mathrm{sin}\,\theta$, and we assume the magnitude of the $\ket{00}$, $\ket{01}$, $\ket{10}$ and $\ket{11}$ terms in the measured state are equal, as is the case during the preparation of cluster states.

\begin{figure}[!htb]
\includegraphics[width=0.7\columnwidth]{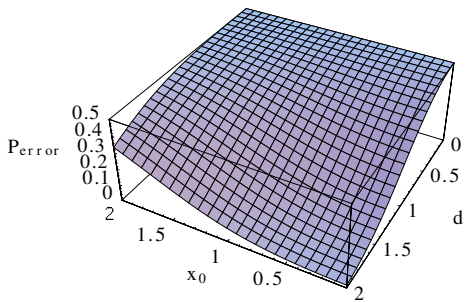}
\caption{Error probability against window width $x_0$, and $d=2\alpha\,\mathrm{sin}\,\theta$.} \label{fig:windowing}
\end{figure}

One caveat is that we have assumed we post-select onto $p=0$. This would give a success probability of 0. Thus, in practise one would post-select onto a small window around $p=0$, between say $\pm p_0$. In this case the effective error rate is obtained by integrating the undesired $\gamma$ terms over the post-selection window. This gives the expression for the error probability
\begin{equation}
P_\mathrm{error} = \frac{\mathrm{erf}[x_0+d] + \mathrm{erf}[x_0 - d]}{2 \, \mathrm{erf}[x_0] + \mathrm{erf}[x_0 + d] + \mathrm{erf}[x_0 - d]},
\end{equation}
where $x_0$ is the width of the post-selection window. In Fig. \ref{fig:windowing} we plot the error probability as a function of window size and $d$. For narrow post-selection windows Eq. \ref{eq:post_selection_error} provides a good approximation to this. Thus, for simplicity the calculations in subsequent sections we will be based on Eq. \ref{eq:post_selection_error}, rather than the more complicated integral form.
Notice that for small $d$ we project into the wrong parity subspace half the time, since here we effectively implement the identity operation. On the other hand, for large $d$ the terms in the probe superposition become almost orthogonal, making the projection closely approximate an even parity projection.

Note that in principle the effect of post-selection noise can be made arbitrarily small by using large $d$. This can be achieved using a strong probe beam ($\alpha\gg 1$) and strong non-linear couplings. However, this is limited by the fact that decoherence associated with bus loss grows with $\alpha \theta$, as will be discussed in Section~\ref{sec:bus_loss}, and also that coupling strengths are experimentally limited.

\section{Bus loss} \label{sec:bus_loss}
\begin{figure}[!htb]
\includegraphics[width=0.7\columnwidth]{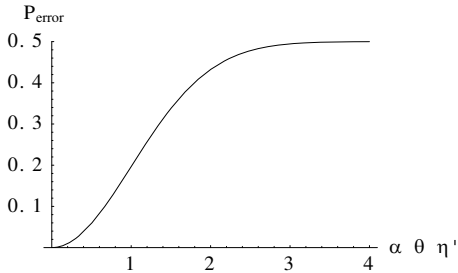}
\caption{Dephasing rate against $\alpha \theta \eta'$.} \label{fig:bus_loss}
\end{figure}

Next we consider the effects of loss in the bus mode. We consider loss between the first and second interactions. We do not need to consider loss before the first interaction, since at this stage the bus is in a coherent state, which does not dephase under loss. Similarly, after the second interaction, the two terms which are post-selected are of the form $\ket{\alpha}\ket{00}+\ket{\alpha}\ket{11}$, which also does not decohere under loss. Following the first non-linear interaction we have
\begin{eqnarray}
\ket{\psi_1} &=& c_{00}\ket\alpha\ket{00} + c_{01}\ket\alpha\ket{01}\nonumber\\
&+& c_{10}\ket{\alpha e^{i\theta}}\ket{10} + c_{11}\ket{\alpha e^{i\theta}}\ket{11}.
\end{eqnarray}
We model loss discretely using a beamsplitter with transmissivity $\eta^2$. The reflected (i.e. loss) mode is then discarded. Applying loss ${\eta'}^2$ (where $\eta' = \sqrt{1-{\eta}^2}$) to the probe mode we obtain
\begin{eqnarray}
\ket{\psi_2} &=& c_{00}\ket{\eta\alpha}_T\ket{\eta'\alpha}_{L}\ket{00}\nonumber\\
&+& c_{01}\ket{\eta\alpha}_T\ket{\eta'\alpha}_{L}\ket{01}\nonumber\\
&+& c_{10}\ket{\eta\alpha e^{i\theta}}_T\ket{\eta'\alpha e^{i\theta}}_{L}\ket{10}\nonumber\\
&+& c_{11}\ket{\eta\alpha e^{i\theta}}_T\ket{\eta'\alpha e^{i\theta}}_{L}\ket{11},
\end{eqnarray}
where $T$ denotes the transmitted mode, and $L$ the loss mode. Here we have used the beamsplitter identity $\hat{U}_{BS}(\eta)\ket{\alpha}_T\ket{0}_L \to \ket{\sqrt\eta\alpha}_T\ket{\sqrt{1-\eta^2}\alpha}_L$. Applying the second interaction and discarding terms that will later be postselected out \footnote{Note that in doing so we implicitly assume are in the regime where $\langle \alpha | \alpha e^{i\theta} \rangle\approx 0$.}, we obtain
\begin{equation}
\ket{\psi_3} = c_{00} \ket{\eta\alpha}_T \ket{\eta'\alpha}_L \ket{00} + c_{11} \ket{\eta\alpha}_T \ket{\eta'\alpha e^{i\theta}}_L \ket{11}.
\end{equation}
Finally we trace out the loss modes $L$ to obtain the output state,
\begin{eqnarray} \label{eq:output_loss}
\hat\rho &=& \mathrm{tr}_L\left(\ket{\psi_3}\bra{\psi_3}\right)\nonumber\\
&=& |c_{00}|^2 \ket{00}\bra{00} + \gamma c_{00}{c_{11}}^* \ket{00}\bra{11}\nonumber\\
&+& \gamma^* {c_{00}}^*c_{11}\ket{11}\bra{00} + |c_{11}|^2\ket{11}\bra{11},
\end{eqnarray}
where
\begin{equation}
|\gamma| = |\langle \eta' \alpha | \eta' \alpha e^{i\theta} \rangle | = \mathrm{exp}\left[\alpha^2 {\eta'}^2 (\mathrm{cos}\,\theta - 1)\right]
\end{equation}
characterizes the decoherence. For small $\theta$ this reduces to
\begin{equation} \label{eq:approx_loss}
|\gamma| = \mathrm{exp}\left[-\frac{1}{2}\alpha^2\theta^2{\eta'}^2\right]=\mathrm{exp}\left[-\frac{1}{8}d^2{\eta'}^2\right].
\end{equation}
Here we have assumed that we post-select onto a narrow window around $p=0$, thereby ignoring post-selecting errors. There are two effects taking place in Eq.~\ref{eq:output_loss}. First, the magnitude of $\gamma$ determines the degree of coherence between the $\ket{00}$ and $\ket{11}$ terms. The phase of $\gamma$ corresponds to a local phase rotation. We assume the local phase rotation can be corrected for. Thus, Eq.~\ref{eq:output_loss} can be described as the ideal measurement process followed by a dephasing channel of the form,
\begin{equation}
\mathcal{E}(\hat\rho) = (1-p)\hat\rho + p\hat{Z}\hat\rho\hat{Z},
\end{equation}
where the dephasing probability is $p = (1-|\gamma|)/2$. Let us consider the general behavior of this expression. First, the dephasing rate increases exponentially with the strength of the coherent probe, asymptotically approaching $p=0.5$ (a perfect dephasing channel) for large $\alpha$.  Second, the dephasing rate is related to the separation of the different components in phase-space, $\theta$. In the limit of no separation ($\theta=0$) there is no dephasing, but clearly the scheme doesn't work at all. The dephasing rate is plotted in Fig.~\ref{fig:bus_loss}.

For comparison, let us consider how this rate relates to estimated fault tolerant thresholds for cluster state quantum computing. Let us assume a fault-tolerant threshold of 1\%. Solving Eq. \ref{eq:approx_loss} for $P\leq 0.01$ we obtain $\alpha \theta \eta' \leq 0.2$. Notice that there is a direct tradeoff between $\alpha$ and $\eta'$. If we have a high loss rate we must use small magnitude coherent states, and vice versa. Also notice that this result provides a tradeoff against Fig. \ref{fig:windowing}. On one hand, Fig. \ref{fig:windowing} requires that we have large $\alpha$ to overcome post-selection error. On the other hand, having large $\alpha$ makes us extremely susceptible to loss. As an example, to suppress post-selection errors to $10^{-2}$ requires $\alpha$ on the order of 110 (assuming $\theta=0.01$). With this value of $\alpha$, suppressing dephasing noise associated with loss to the same level requires a loss rate of roughly 3\%. Notice that for smaller $\theta$ we can tolerate higher loss rates. However, one must keep in mind that the flip-side of this high loss tolerance is that for small $\theta$ one must post-select on narrower windows, which reduces the gate's success probability. Thus, loss tolerance is effectively boosted via post-selection.

In our analysis we have considered a discrete channel loss model, where loss occurs at one location between the two interactions. A continuous loss model within the non-linear medium was considered in Ref. \cite{bib:Jeong06} for the same parity gate.

\section{Mismatched coupling rates} \label{sec:mis_coupling}
In an ideal implementation of the parity gate we choose $\theta_A=-\theta_B$. This way a postselection onto $p\approx 0$ cannot distinguish between the $\ket{00}$ and $\ket{11}$ logical states, leaving us with a coherent projection onto the subspace spanned by these two vectors -- the even parity subspace. In practise, imperfect calibration and dynamic changes in the non-linear interaction strengths imply that there will be some small mismatch between $\theta_A$ and $\theta_B$. There are two distinct ways in which this effect may arise, \emph{known} and \emph{unknown} mismatch. Known mismatches result in a well defined unitary operations being applied, which can be either subsequently undone or tolerated since they are known. Unknown mismatch on the other hand results in integration over different possible unitary operations, resulting in a mixing effect. Clearly this effect is far more poignant. Thus, in this section we consider the latter effect.

Beginning with the general input state
\begin{equation}
\ket{\psi_\mathrm{in}} = \ket\alpha(c_{00}\ket{00} + c_{01}\ket{01} + c_{10}\ket{10} + c_{11}\ket{11}),
\end{equation}
we apply two consecutive interactions with distinct coupling stengths, $\hat{U}_{PB}(-\theta_B)\hat{U}_{PA}(\theta_A)$. The output state is
\begin{eqnarray}
\ket{\psi_\mathrm{out}} &=& \hat{U}_{PB}(-\theta_B) \hat{U}_{PA}(\theta_A) \ket{\psi_\mathrm{in}} \nonumber\\
&=& c_{00} \ket\alpha \ket{00} + c_{01} \ket{\alpha e^{-i\theta_B}} \ket{01} + \nonumber\\
&+& c_{10 }\ket{\alpha e^{i\theta_A}} \ket{10} + c_{11} \ket{\alpha e^{i(\theta_A-\theta_B)}} \ket{11}.
\end{eqnarray}
Next we post-select onto a $p$-quadrature measurement in a window around $p\approx 0$. Upon post-selection we obtain
\begin{eqnarray}
\ket{\psi_\mathrm{cond}} &=& c_{00}\gamma_0\ket{00} + c_{01}\gamma_{\theta_B}\ket{01} \nonumber\\
&+& c_{10}\gamma_{\theta_A}\ket{10} + c_{11}\gamma_{\theta_A-\theta_B}\ket{11}.
\end{eqnarray}
Ideally we aim to adjust $\theta_A$ and $\theta_B$ such that $\theta_A=\theta_B$ and $\gamma_{\theta_A}\approx\gamma_{\theta_A}\approx 0$. In this case the expression reduces to the ideal result,
\begin{equation}
\ket{\psi_\mathrm{cond}} = \gamma_0 (c_{00}\ket{00} + c_{11}\ket{11}),
\end{equation}
where the leading factor of $\gamma_0$ is associated with the non-determinism of the measurement.

In the case where the coupling strengths are not perfectly matched, $\Delta = \theta_A - \theta_B \neq 0$, we have the more general expression
\begin{equation}
\ket{\psi_\mathrm{cond}} = \gamma_0 c_{00}\ket{00} + \gamma_\Delta c_{11}\ket{11},
\end{equation}
where we again ignore post-selection errors by assuming $\gamma_{\theta_A}\approx\gamma_{\theta_B}\approx 0$. If $\Delta$ is known this gives us a biased parity projection. In principle, preparation of the known state $\ket{\psi_\mathrm{in}}=\ket{+}\ket{+}$ followed by quantum state tomography (QST) \cite{bib:NielsenChuang00} allows us to infer $\Delta$, from which coupling strengths may be calibrated. However, in practise there will always be some uncertainty in $\Delta$ due to dynamic changes and the precision of the calibration procedure. So we next consider the case where there is some variance in $\Delta$. We assume $\Delta$ ranges between $\pm\Delta_0$. First rewrite
\begin{eqnarray}
\hat\rho_\mathrm{cond} &=& |\gamma_0|^2 (|c_{00}|^2 \ket{00}\bra{00} + {\lambda_\Delta}^* c_{00}{c_{11}}^* \ket{00}\bra{11} \nonumber\\
&+& {\lambda_\Delta} {c_{00}}^*c_{11} \ket{11}\bra{00} + |{\lambda_\Delta}|^2 |c_{11}|^2 \ket{11}\bra{11}), \nonumber\\
\end{eqnarray}
where $\lambda_\Delta = \gamma_\Delta / \gamma_0$. Summing over $\Delta$ in the range $\pm\Delta_0$ we obtain
\begin{equation}
\hat\rho = \frac{1}{2\Delta_0} \int_{-\Delta_0}^{\Delta_0} \hat\rho_\mathrm{cond}\,\mathrm{d}\Delta.
\end{equation}
Here we have assumed that that all $\hat\rho_\mathrm{cond}$ within the window $\pm\Delta_0$ are equally likely. As before, $\hat\rho$ can be regarded as an ideal measurement process followed by a dephasing channel of the form,
\begin{equation}
\hat\rho = \mathcal{E}(\ket{\psi_\mathrm{b}}\bra{\psi_\mathrm{b}}) = (1-p)\ket{\psi_\mathrm{b}}\bra{\psi_\mathrm{b}} + p\hat{Z}\ket{\psi_\mathrm{b}}\bra{\psi_\mathrm{b}}\hat{Z}.
\end{equation}
$\hat{Z}$ can act on either qubit (since this is an even-parity state, it is exchange symmetric), and the dephasing probability $p$ is given by
\begin{equation}
p = \frac{1}{2}\left[1 - \frac{\int_{-\Delta_0}^{\Delta_0} \lambda_\Delta \,\mathrm{d}\Delta}{\sqrt{2\Delta_0 \int_{-\Delta_0}^{\Delta_0} |\lambda_\Delta|^2 \,\mathrm{d}\Delta}} \right].
\end{equation}
$\ket{\psi_\mathrm{b}}$ is the biased parity projected state,
\begin{equation} \label{eq:biased_state}
\ket{\psi_\mathrm{b}} = |\gamma_0|^2 \left(c_{00}\ket{00} + \mu c_{11} \ket{11}\right),
\end{equation}
where the bias in the projection is determined by the parameter,
\begin{equation}
\mu = \frac{1}{2\Delta_0} \sqrt{\int_{-\Delta_0}^{\Delta_0} |\lambda_\Delta|^2 \,\mathrm{d}\Delta}.
\end{equation}
Note that while the later effect, biased projection, is undesired, it is a coherent process and therefor might be tolerated in a variety of situations. For example, Campbell et al. \cite{bib:Campbell07} considered the construction of cluster states in the presence of biased projections. The dephasing effect on the other hand is much for troublesome and represents a true decohering effect that cannot be undone without error correction. As an example, if we wish to suppress the dephasing probability to below 1\%, the uncertainty in the coupling mismatch is bounded by $\Delta_0<0.64$. Thus the scheme is quite resilient against unknown coupling rate mismatch. However, for this value of $\Delta_0$ the bias is given by $\mu = 0.66$, a heavily biased state.

\begin{figure}
\includegraphics[width=0.7\columnwidth]{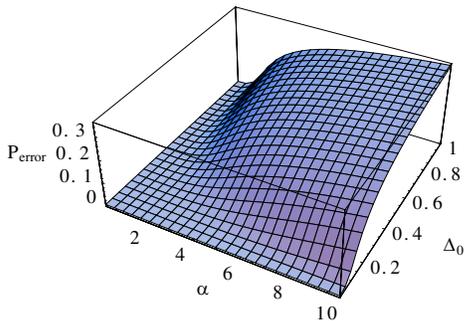}
\caption{Error probability against probe strength and the variance in the coupling mismatch.} \label{fig:mismatched_coupling}
\end{figure}

In Fig.~\ref{fig:mismatched_coupling} we plot the dephasing probability against probe strength and the variance in the coupling mismatch.


\section{Mode-mismatch} \label{sec:mode_mismatch}
In optical implementations of the parity gate, an additional practical consideration is mode-mismatch. This corresponds to imperfect spatio-temporal overlap of the photonic qubit and bus modes during the respective interaction. Mode-mismatch has previously been considered in the context of LOQC protocols in Refs. \cite{bib:RohdeRalph05, bib:RohdeRalphMunro06, bib:RohdeRalph06}. In this section we consider how this affects gate operation. We model mode-mismatch using the eigenmode decomposition technique introduced in Ref.~\cite{bib:RohdeMauererSilberhorn07}.

Assume the coherent probe is characterized by a spatio-temporal mode-function $\phi$, and the photonic qubits by mode-functions $\psi_A$ and $\psi_B$ respectively. A single photon state with mode-function $\psi$ can be decomposed into components overlapping with and orthogonal to the probe beam,
\begin{equation}
\ket{1}_\psi = \lambda_1\ket{1}_\phi\ket{0}_{\bar\phi} + \lambda_0\ket{0}_\phi\ket{1}_{\bar\phi},
\end{equation}
where $\ket{1}_\psi$ represents the single photon state characterized by mode-function $\psi$, and $\bar\phi$ is the component of $\psi$ orthogonal to $\phi$. $\lambda_1$ represents the degree of mode overlap, and varies between 0 and 1. It can be calculated as
\begin{eqnarray}
\lambda_1 &=& \int \phi(\vec\omega)^*\psi(\vec\omega)\,\mathrm{d}\vec\omega,\nonumber\\
\lambda_0 &=& \sqrt{1-{\lambda_1}^2},
\end{eqnarray}
where $\vec\omega$ are the spatio-temporal degrees of freedom. Using this decomposition we can redefine the action of the interaction operator $\hat{U}$,
\begin{eqnarray}
\hat{U}(\theta)\ket\alpha\ket{0} &=& \ket\alpha\ket{0} \nonumber\\
\hat{U}(\theta)\ket\alpha\ket{1} &=& \lambda_1 \ket{\alpha e^{i\theta}}\ket{1}_\phi\ket{0}_{\bar\phi} + \lambda_0 \ket{\alpha}\ket{0}_\phi\ket{1}_{\bar\phi}.
\end{eqnarray}
Thus, when a single photon is present, only the component overlapping with the coherent probe undergoes any interaction. The non-overlapping component is unaffected.

\begin{figure}[!htb]
\includegraphics[width=0.7\columnwidth]{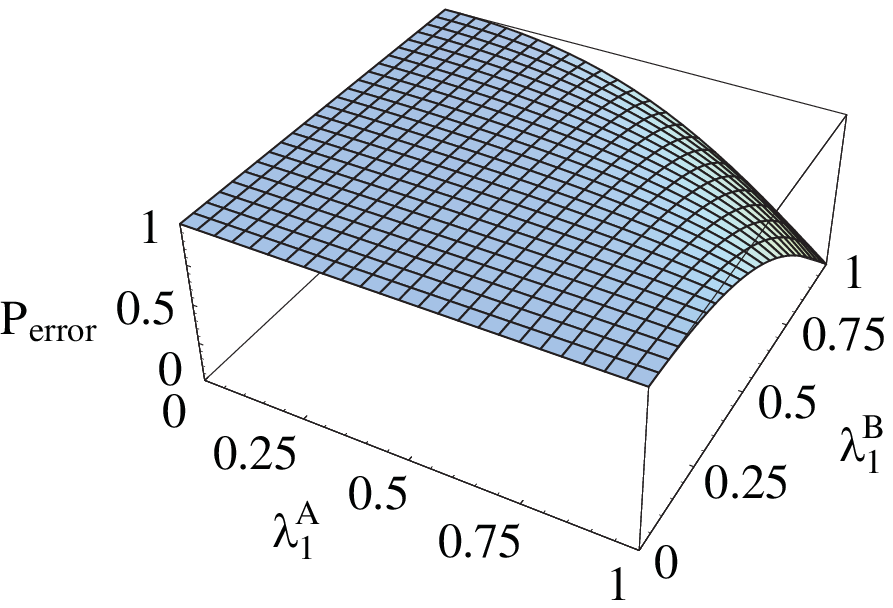}
\includegraphics[width=0.75\columnwidth]{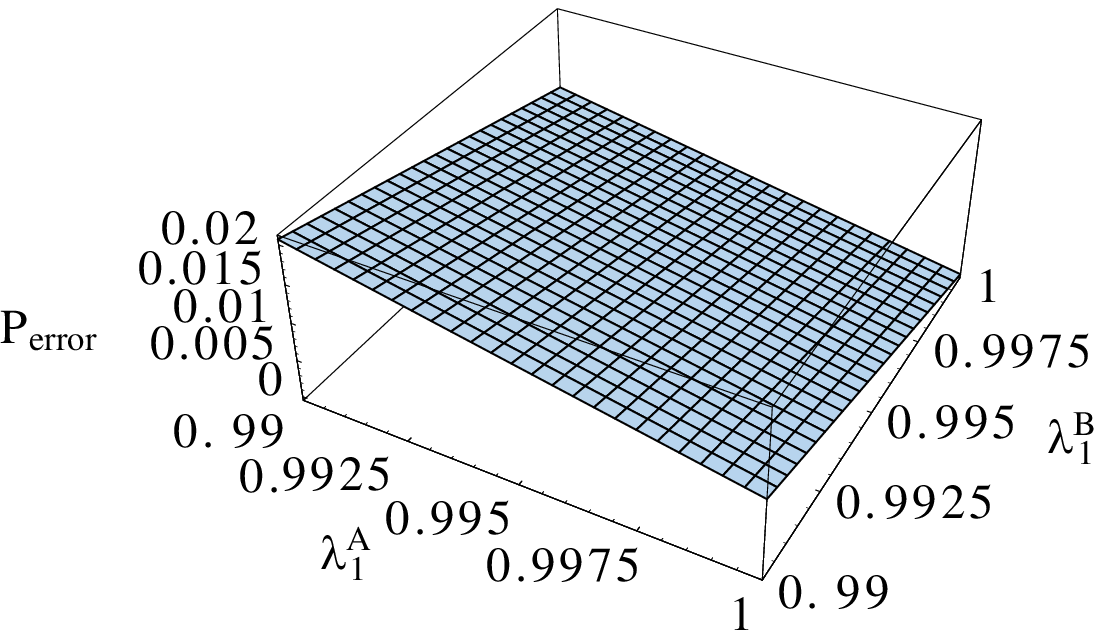}
\caption{Measurement error probability against the two mode-matching parameters.} \label{fig:mode_mismtach}
\end{figure}

Applying this decomposition to the input state and applying the interactions and projection, we obtain
\begin{eqnarray}
\ket{\psi_\mathrm{out}} &=& c_{00}\gamma_0\ket{00}\nonumber\\
&+& c_{01}[\lambda_1^B \gamma_{-\theta_B}\ket{01_\phi} + \lambda_0^B \gamma_0\ket{01_{\bar\phi}}]\nonumber\\
&+& c_{10}[\lambda_1^A \gamma_{\theta_A}\ket{1_\phi 0} + \lambda_0^A \gamma_0\ket{1_{\bar\phi}0}]\nonumber\\
&+& c_{11}[\lambda_1^A\lambda_1^B \gamma_\Delta \ket{1_\phi 1_\phi} + \lambda_1^A\lambda_0^B \gamma_{\theta_A}\ket{1_\phi 1_{\bar\phi}}\nonumber\\
&+& \lambda_0^A\lambda_1^B \gamma_{-\theta_B}\ket{1_{\bar\phi} 1_\phi} + \lambda_0^A\lambda_0^B \gamma_0 \ket{1_{\bar\phi}1_{\bar\phi}}].
\end{eqnarray}
Consider the usual limit where $\theta_A=-\theta_B$ is sufficiently large such that $\gamma_{\theta_A}=\gamma_{\theta_B}\approx 0$. Now the output state reduces to
\begin{eqnarray}
\ket{\psi_\mathrm{out}} &=& c_{00}\ket{00} + c_{01}\lambda_0^B\ket{01_{\bar\phi}} + c_{10}\lambda_0^A\ket{1_{\bar\phi}0}\nonumber\\
&+& c_{11}[\lambda_1^A\lambda_1^B\ket{1_\phi 1_\phi} + \lambda_0^A\lambda_0^B\ket{1_{\bar\phi}1_{\bar\phi}}],
\end{eqnarray}
up to normalization. In this case our projection is no longer a perfect parity projection, but rather described by the more general projection operator
\begin{eqnarray}
\hat\Lambda &=& \ket{00}\bra{00} + \lambda_0^B \ket{01}_{\bar\phi}\bra{01}_{\bar\phi} + \lambda_0^A \ket{10}_{\bar\phi}\bra{10}_{\bar\phi}\nonumber\\
&+& \lambda_0^A \lambda_0^B \ket{11}_{\bar\phi}\bra{11}_{\bar\phi} + \lambda_1^A\lambda_1^B \ket{11}_\phi\bra{11}_\phi.
\end{eqnarray}
Note that in the ideal case where we have perfect mode overlap, $\lambda_1^{A,B}=1$, in which case this operator reduces to the ideal parity projection operator. In the general case we can regroup the projector as
\begin{eqnarray}
\hat\Lambda &=& \lambda_1^A\lambda_1^B (\ket{00}\bra{00} + \ket{11}_\phi\bra{11}_\phi)\nonumber\\
&+& (1-\lambda_1^A\lambda_1^B)\ket{00}\bra{00} + \lambda_0^B \ket{01}_{\bar\phi}\bra{01}_{\bar\phi} \nonumber\\
&+& \lambda_0^A \ket{10}_{\bar\phi}\bra{10}_{\bar\phi} + \lambda_0^A \lambda_0^B \ket{11}_{\bar\phi}\bra{11}_{\bar\phi}.
\end{eqnarray}
whereby the first term represents the ideal parity projection operator in the desired spatio-temporal optical mode. As before, the measurement error probability is given by the relative magnitude of the undesired terms,
\begin{widetext}
\begin{equation}
P_\mathrm{error} = 1 - \frac{2|\lambda_1^A\lambda_1^B|^2}{2|\lambda_1^A\lambda_1^B|^2 + |1 - \lambda_1^A\lambda_1^B|^2 + |\lambda_0^A|^2 + |\lambda_0^B|^2 + |\lambda_0^A\lambda_0^B|^2}.
\end{equation}
\end{widetext}
Measurement error probability is shown in Figure~\ref{fig:mode_mismtach}, plotted against the two mode-matching parameters $\lambda_1^A$ and $\lambda_1^B$.
As mode-overlap approaches unity in both interactions the error probability drops to zero, the expected ideal-case result. As mode overlap deteriorates, $\lambda_1^A<1$ and $\lambda_1^B<1$, the probe beam ceases to interact with the photonic qubits. Thus, the probe beam will not pick up any conditional phase shifts, leaving it in the $\ket\alpha$ state, separable from the qubit states. Thus, the qubits undergo the identity operation. Note that in the limit of complete mismatch the error rate goes to 1 instead of 1/2 as for the other error channels. This is because the error channel comprises two distinct types of error. First, in the qubit space, the qubits have undergone the identity operation rather than the parity operation. Second, the qubits do not reside in the desired spatio-temporal mode. We define the error probability as the probability that the state ends up in the undesired state -- either in terms of its logical qubit value, or its spatio-temporal state. It is the later effect that pushes the error rate to unity for complete mode-mismatch.

Note that following the parity projection, the photonic qubits are projected into a spatio-temporal mode-structure different to the original one. Thus, if such measurements are cascaded the $\lambda$ coefficients will need to be re-calculated for each parity measurement. Therefore, the above analysis applies only to a single application of the gate.


In a recent paper by Shapiro et al. \cite{bib:Shapiro07} they consider a parity gate based on single photons and an intense probe beam. All the photons and probe beams are assumed to be pulses in a propagating system. In this case it was found that there was significant extra phase noise from the pulses not being single mode. However in other physical embodiments (for instance cavity QED) this phase noise can be significantly minimized \cite{bib:Milburn07, bib:Ladd07}.

\section{Self-Kerr interactions} \label{sec:self_kerr}

The weak cross-Kerr interaction required for the implementation of the parity is somewhat idealized. In practise the Hamiltonian describing such an interaction will also include a self-Kerr term of the form
\begin{equation}
\hat{H}_\mathrm{SK} = \lambda{\hat{n}_p}^2.
\end{equation}
When acting on logical qubits this term simply results in a well-defined phase rotation, which can easily be undone. We therefore restrict ourselves to considering the effect on the bus mode where the resulting evolution is not so trivial and results in squeezing of the coherent probe. Let us label the unitary evolution corresponding to this self-Kerr term in the Hamiltonian as $\hat{U}_\mathrm{sk}$. To analyze the effect this has on the parity gate note that the self-Kerr term commutes with the desired cross-Kerr term. Therefore, we can commute the self Kerr terms to the end of the circuit. Then the circuit can be treated as the ideal parity circuit, followed by the self-Kerr operations, followed by the projective measurement. Recall that the state following the two weak non-linear interactions, but prior to measurement, is of the form
\begin{eqnarray}
\ket{\psi_\mathrm{out}} &=& c_{00}\ket\alpha\ket{00} + c_{01}\ket{\alpha e^{i\theta}}\ket{01}\nonumber\\
&+& c_{10}\ket{\alpha e^{-i\theta}}\ket{10} + c_{11}\ket\alpha\ket{11}.
\end{eqnarray}
Next we apply the two self-Kerr terms that we have commuted to the end, which we jointly label $\hat{U}_\mathrm{sk}' = \hat{U}_\mathrm{sk,1}\hat{U}_\mathrm{sk,2}$, and apply the $p$-measurement projector to obtain
\begin{eqnarray}
\ket{\psi_\mathrm{cond}} &=& c_{00}\bra{p}\hat{U}_\mathrm{sk}'\ket\alpha\ket{00} + c_{01}\bra{p}\hat{U}_\mathrm{sk}'\ket{\alpha e^{i\theta}}\ket{01}\nonumber\\
&+& c_{10}\bra{p}\hat{U}_\mathrm{sk}'\ket{\alpha e^{-i\theta}}\ket{10} + c_{11}\bra{p}\hat{U}_\mathrm{sk}'\ket\alpha\ket{11}.\nonumber\\
\end{eqnarray}
Let us redefine a new $\gamma$ parameter as
\begin{eqnarray}
\gamma_{p,\theta}' &=& \bra{p}\hat{U}_\mathrm{sk}'\ket{\alpha e^{i\theta}}\nonumber\\
&=& \pi^{-1/4}e^{-\frac{1}{2}p^2-|\alpha|^2/2} \sum_n \frac{(\alpha e^{i\theta})^n e^{i\lambda n^2}e^{-in\pi/2}}{2^{n/2}n!}H_n(p) \nonumber\\
\end{eqnarray}
Then the output state is simply given by
\begin{eqnarray}
\ket{\psi_\mathrm{cond}} &=& c_{00}\gamma_{p,0}'\ket{00} + c_{01}\gamma_{p,\theta}'\ket{01}\nonumber\\
&+& c_{10}\gamma_{p,-\theta}'\ket{10} + c_{11}\gamma_{p,0}'\ket{11}.
\end{eqnarray}
This output state is of the same form as for the ideal parity gate, but with different $\gamma$ coefficients. As before, we can relate the form of this state to a two-qubit error model by grouping this expression into ideal and error terms. Then the error probability is given by the relative magnitude of these terms. Specifically,
\begin{equation}
P_\mathrm{error} = \frac{|\gamma_{p,\theta}'|^2 + |\gamma_{p,-\theta}'|^2}{2|\gamma_{p,0}'|^2 + |\gamma_{p,\theta}'|^2 + |\gamma_{p,-\theta}'|^2}.
\end{equation}

\begin{figure}[!htb]
\includegraphics[width=0.8\columnwidth]{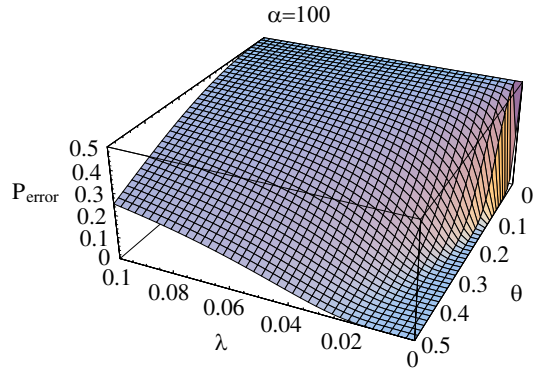}
\caption{Error probability against the strength of the self-Kerr term $\lambda$ and cross-Kerr term $\theta$.} \label{fig:P_alpha_lambda}
\end{figure}

In Fig.~\ref{fig:P_alpha_lambda} we plot the error probability $P_\mathrm{error}$ the self-Kerr strength $\lambda$ and cross-Kerr strength $\theta$ for two different probe strengths $\alpha$. There are two important observations here. Firstly, $\lambda$ must be kept very small compared to the the cross-Kerr strength to suppress errors. Secondly, the larger the self-Kerr strength the larger the cross-Kerr strength must be to maintain the same error rate. 
Qualitatively, the mechanism by which the self-Kerr effect causes gate error is as follows. The self Kerr effect induces squeezing on the probe beam, causing it to be circularly `smeared out' in phase space. Because the terms in the superposition are smeared the $P$-quadrature measurements become ambiguous, resulting in uncertainty as to which terms in the superposition were measured.

\section{Conclusion} \label{sec:conclusion}
We have considered the effects of several dominant sources of imperfection is the bus-mediated parity gate. Specifically, we analyzed post-selection errors, bus loss, mismatched coupling rates, mode-mismatch and self-Kerr interactions. For each of these effects we derived error models characterizing the system's performance. This is a useful approach since error rates in standard error models can be directly related to relevant fault-tolerance thresholds. In general these effects all degrade performance and place stringent requirements if such a scheme is to be operated in a low error regime.

The bus mediated parity gate effectively maps loss in the bus to dephasing noise on the logical qubits. It is known from fault-tolerant LOQC simulations \cite{bib:Dawson05, bib:Dawson06} that the threshold for loss is much higher than for dephasing noise. This arises because loss is a located error in LOQC, whereas dephasing is unlocated. This is unfortunate for the bus mediated gate, since it maps an error with higher threshold to one with a lower threshold. Thus we expect the bus mediated gate to be less tolerant against loss than the LOQC gate. On the other hand, a cluster state construction with the bus mediated gate it likely to be significantly more resource efficient than the comparable LOQC parity gate. This is because the bus mediated gate is non-destructive, unlike LOQC gates which destroy a least one qubit upon success. Interestingly, a similar effect occurs in the opposite direction when we consider post-selection errors. Ordinarily we post-select on a window around $p=0$. There will be dephasing at a rate related to the width of the window, due to post-selection errors. These dephasing errors can be minimized by narrowing the window, which comes at the expense of gate success probability. Thus, by narrowing the window we effectively map the less desirable dephasing errors to heralded errors. Put simply, in order to minimize post-selection errors one must use large $d=2\alpha\,\mathrm{sin}\,\theta$. On the other hand, large $d$ makes us susceptible to loss. Thus, loss requirements are extremely tight.

In the case of mismatched coupling rates we first recognized that known mismatched coupling rates lead to known unitary operations, which in principle can be corrected for. We then turned our attention to the case where there is some uncertainty in the coupling rates. This results in two distinct effects. Firstly, dephasing occurs. This comes about as a result of mixing over the possible range of coupling rates. Again, dephasing is a standard error model to which known error correcting protocols can be applied. The second effect was a bias in the parity projection. That is, rather than applying the projection $\ket{00}\bra{00}+\ket{11}\bra{11}$, the biased projection $\ket{00}\bra{00}+\mu\ket{11}\bra{11}$ is applied.

Next we considered the effects of mode-mismatch, a result that applies only to the photonic implementation of the scheme. Here the resulting projection could not be described by a trivial error model. Rather the measurement projector contained many terms. However, the projector can be factored into desired and undesired terms, allowing for the definition of an error probability. Our results indicate that in order to suppress the error probability to below 1\% requires mode-overlap on the order of $0.995$.

Finally we considered the effects of self-Kerr interactions. This effect results in a measurement projector containing undesired odd parity terms. Importantly, this does not result in mixing. Rather it result in a coherent projection with some undesired terms. Because no mixing takes place, this type of error might be more easily tolerated in a variety of situations.

In summary, our results indicate that, while possible in principle, the requirements for implementing bus-mediated quantum computing are very demanding. Our results are of particular relevance to cluster state quantum computing, where gates such as the one considered here are directly applicable.

\begin{acknowledgments}
We thank Nick Menicucci for helpful discussions. PR acknowledges support from the Australian Research Council, Queensland State Government, and DTO-funded U.S. Army Research Office Contract No. W911NF-05-0397. BM and KN acknowledge support from MEXT and QAP.
\end{acknowledgments}

\bibliography{paper}

\end{document}